\documentclass[%
 aip,
 amsmath,amssymb,
 reprint,%
]{revtex4-1}

\usepackage{graphicx}
\usepackage{dcolumn}
\usepackage{bm}

\usepackage[utf8]{inputenc}
\usepackage[T1]{fontenc}
\usepackage{mathptmx}
\usepackage{etoolbox}
\usepackage{xcolor}
\usepackage{caption}
\usepackage{algorithm}
\usepackage{algpseudocode}
\usepackage[normalem]{ulem}
\usepackage{booktabs}
\newcommand{\R}{\mathbb R}

\definecolor{fpgreen}{rgb}{0,.5,0}
\definecolor{rablue}{rgb}{0,0,255}

\makeatletter
\def\@email#1#2{%
 \endgroup
 \patchcmd{\titleblock@produce}
  {\frontmatter@RRAPformat}
  {\frontmatter@RRAPformat{\produce@RRAP{*#1\href{mailto:#2}{#2}}}\frontmatter@RRAPformat}
  {}{}
}%
\makeatother
\begin{document}

\preprint{AIP/123-QED}

\title[CASSCF response equations revisited]{CASSCF response equations revisited: a simple and efficient iterative algorithm}
\author{Riccardo Alessandro}
\author{Ivan Giann\'i}
\author{Federica Pes}
\author{Tommaso Nottoli}
\email{tommaso.nottoli@phd.unipi.it}
\author{Filippo Lipparini}
\email{filippo.lipparini@unipi.it}
\affiliation{ 
Dipartimento di Chimica e Chimica Industriale, Universit\`a di Pisa\\
Via G. Moruzzi 13, 56124 Pisa, Italy}%

\date{\today}

\begin{abstract}
We present an algorithm to solve the CASSCF linear response equations that is both simple and efficient. The algorithm makes use of the well established symmetric and antisymmetric combinations of trial vectors, but further orthogonalizes them with respect to the scalar product induced by the response matrix. This leads to a standard, symmetric, block eigenvalue problem in the expansion subspace that can be solved by diagonalizing a symmetric, positive definite matrix half the size of the expansion space. Preliminary numerical tests show that the algorithm is robust and stable.
\end{abstract}
\maketitle

Linear response calculations are commonly encountered in computational chemistry, as they are used to compute excitation energies and transition properties. The linear response equations for CASSCF have the general form, assuming that real basis functions are used:\cite{norman2018principles, yeager1979multiconfigurational, yeager1981evaluation, sasagane1990multiconfiguration, christiansen1998response, helgaker2012recent, norman2011perspective}
\begin{equation}
    \begin{pmatrix}
        A & B \\
        B & A
    \end{pmatrix}
    \begin{pmatrix}
        \bm{y} \\ \bm{z}
    \end{pmatrix}
    = 
    \omega
    \begin{pmatrix}
        \Sigma & \Delta \\
        -\Delta & -\Sigma
    \end{pmatrix}
    \begin{pmatrix}
        \bm{y} \\ \bm{z}
    \end{pmatrix},
    \label{eq:CASGenEig}
\end{equation}
where
$A$, $B$, and $\Sigma$ are symmetric matrices in $\mathbb{R}^{n\times n}$, $\Delta$ is an antisymmetric matrix in $\mathbb{R}^{n\times n}$, and $\bm{y}, \bm{z}$ are vectors in $\mathbb{R}^n$. Here $n$ represents the size of the problem, e.g., the number of occupied times number of virtual orbitals for Hartree-Fock (HF) and Kohn-Sham Density Functional Theory (KS-DFT), and we further assume that $\Sigma$ is positive definite, as are the combinations $A+B$ and $A-B$, conditions that are met if the ground state solution is stable. The eigenvalue problem \eqref{eq:CASGenEig} has been investigated in the standard form as well as in the generalized one: in \cite{bai2012minimization, bai2013minimization, bai2014minimization} the authors developed a minimization principle to find a few smallest eigenvalues and the corresponding eigenvectors. For HF and KS-DFT response theory, eq. \eqref{eq:CASGenEig} is simplified, as $\Sigma$ becomes the identity matrix and $\Delta$ the zero matrix. For the sake of brevity, we also write the response equations as
\begin{equation}
    \label{eq:caseigcomp}
    \Lambda \bm{x} = \omega \Omega \bm{x}.
\end{equation}
Eq. \eqref{eq:caseigcomp} is a generalized eigenvalue problem with a non-positive-definite metric $\Omega$, which means that the standard procedure used in other quantum chemical calculations, where either a Cholesky decomposition of the metric or $\Omega^{-1/2}$ are computed in order to transform eq. \eqref{eq:caseigcomp} into a standard eigenvalue problem cannot be used. However, the linear response equations have an important property. If $\{\omega,(\bm{y},\bm{z})^T\}$ is a solution, then $\{-\omega, (\bm{z},\bm{y})^T\}$ is also a solution to the same problem. In other words, the eigenvalues of eq. \eqref{eq:caseigcomp} always appear as positive-negative pairs. It has been suggested in the literature \cite{olsen1988solution, jo1988linear, saue2003linear, helmich2019casscf} that this symmetry should be encoded in the iterative algorithm used to solve the response equations, usually a modified version of Davidson's method,\cite{davidson1975iterative} by expanding each eigenvector as the linear combination of two sets of vectors 
\begin{equation}
    \label{eq:eigenvector}
    \bm{x} = \sum_{i=1}^k \left ( u^+_i \bm{b}_i(+) + u^-_i \bm{b}_i(-) \right ),
\end{equation}
where $u_i^\pm\in\R$ and the expansion vectors are defined as follows:
\begin{equation}
    \label{eq:expansion}
    \bm{b}_i(+) =
    \begin{pmatrix}
        \bm{b}_i^+ \\ \bm{b}_i^+
    \end{pmatrix} \in \mathbb{R}^{2n},
    \quad
    \bm{b}_i(-) =
    \begin{pmatrix}
        \bm{b}_i^- \\ - \bm{b}_i^-
    \end{pmatrix} \in \mathbb{R}^{2n},
\end{equation}
with $\bm{b}_i^+, \bm{b}_i^- \in \mathbb{R}^n$. In the following, we use the notation, for a generic vector $\bm{v} \in \mathbb{R}^{2n}$, $\bm{v}(+) = (\bm{v}^+, \bm{v}^+)^T$ and $\bm{v}(-) = (\bm{v}^-, -\bm{v}^-)^T$, with $\bm{v}^+, \bm{v}^- \in \mathbb{R}^n$ to denote symmetric and antisymmetric vectors, respectively.
To keep the notation not too cumbersome, we only consider here the case where a single eigenvector is searched, as the generalization to many eigenvectors is straightforward.
This choice of expansion space is particularly advantageous, as it gives rise to blocked reduced matrices for problem \eqref{eq:CASGenEig}.
In fact,
\begin{equation}
    \label{eq:MV1}
    \begin{aligned}
    \Lambda \bm{b}(+) &= 
    \begin{pmatrix}
        (A + B)\bm{b}^+ \\
        (A + B)\bm{b}^+
    \end{pmatrix} = 
    \bm{\sigma}(+),\quad \bm{\sigma}^+=(A+B)\bm{b}^+, \\    
    \Lambda \bm{b}(-) &= 
    \begin{pmatrix}
        (A - B)\bm{b}^- \\
        -(A - B)\bm{b}^-
    \end{pmatrix} = 
    \bm{\sigma}(-),\quad \bm{\sigma}^-=(A-B)\bm{b}^-,
    \end{aligned}
\end{equation}
and
\begin{equation}
    \label{eq:MV2}
    \begin{aligned}
    \Omega \bm{b}(+) &= 
    \begin{pmatrix}
        (\Sigma + \Delta)\bm{b}^+ \\
        -(\Sigma + \Delta)\bm{b}^+
    \end{pmatrix} = 
    \bm{\tau}(-),\quad \bm{\tau}^- =(\Sigma+\Delta)\bm{b}^+,\\    
    \Omega \bm{b}(-) &= 
    \begin{pmatrix}
        (\Sigma - \Delta)\bm{b}^- \\
        (\Sigma - \Delta)\bm{b}^-
    \end{pmatrix} = 
    \bm{\tau}(+),\quad \bm{\tau}^+ =(\Sigma-\Delta)\bm{b}^-.
    \end{aligned}
\end{equation}
In the spirit of Davidson's method, the response equations can be therefore solved as follows. 

Let $\mathcal{V}_k^+$ and $\mathcal{V}_k^-$ be the symmetric and antisymmetric expansion subspaces, respectively, whose dimension increases with $k$. Here $k$ represents the number of iterations already performed. 
Let
\begin{equation*}
V_k(+)= \begin{pmatrix} \bm{b}_1(+) & \ldots & \bm{b}_k(+) \end{pmatrix},\quad 
V_k(-)= \begin{pmatrix} \bm{b}_1(-) & \ldots & \bm{b}_k(-) \end{pmatrix}
\end{equation*}
be matrices in $\mathbb{R}^{2n\times k}$ which columns are the symmetric and antisymmetric expansion vectors, respectively.
Let the columns of the matrices $V_k(\pm)$ form an orthonormal basis for the subspaces $\mathcal{V}_k^\pm$.
Let us also introduce the spaces 
\begin{equation*}
\mathcal{LV}_k(\pm)=\{ \Lambda \bm{b}_1(\pm),\ldots, \Lambda \bm{b}_k(\pm)\}=\{ \bm{\sigma}_1(\pm),\ldots,\bm{\sigma}_k(\pm) \}
\end{equation*}
and 
\begin{equation*}
\mathcal{BV}_k(\pm)=\{ \Omega \bm{b}_1(\mp),\ldots, \Omega \bm{b}_k(\mp)\}=\{ \bm{\tau}_1(\pm),\ldots,\bm{\tau}_k(\pm) \}
\end{equation*}
that collect the applications of the matrices $\Lambda$ and $\Omega$ to the expansion vectors.
We remark that in practice these spaces are not completely used in the algorithm, but only the \emph{half}-dimensional spaces. Specifically, we consider the $n\times k$ matrices 
$V_k^\pm = \begin{pmatrix} \bm{b}^\pm_1 & \ldots & \bm{b}^\pm_k \end{pmatrix}$, the spaces $\mathcal{LV}_k^\pm = \{ \bm{\sigma}_1^\pm,\ldots,\bm{\sigma}_k^\pm \}$ and $\mathcal{BV}_k^\pm = \{ \bm{\tau}_1^\pm,\ldots,\bm{\tau}_k^\pm\}$, and the corresponding matrices $LV_k^\pm = \begin{pmatrix} \bm{\sigma}_1^\pm & \ldots & \bm{\sigma}_k^\pm \end{pmatrix} $ and
$BV_k^\pm = \begin{pmatrix} \bm{\tau}_1^\pm & \ldots & \bm{\tau}_k^\pm \end{pmatrix}$.

The Rayleigh-Ritz variational procedure is then performed to compute the $\bm{u}^\pm$ coefficients from eq. \eqref{eq:eigenvector} by solving the projection of eq. \eqref{eq:CASGenEig} into the subspace union of $V^+_k$ and $V^-_k$. We get the $2k\times 2k$ problem
\begin{equation}
    \label{eq:reduced}
    \begin{pmatrix}
        E^+ & 0 \\
        0 & E^- \\
    \end{pmatrix}
    \begin{pmatrix}
        \bm{u}^+ \\ \bm{u}^-
    \end{pmatrix}
    = \omega_{k+1}
    \begin{pmatrix}
        0 & S^T \\
        S & 0 \\
    \end{pmatrix}
    \begin{pmatrix}
        \bm{u}^+ \\ \bm{u}^-
    \end{pmatrix},
\end{equation}
where
\begin{equation}
    \label{eq:redmatE}
    E^\pm = (V^\pm_k)^T LV_k^\pm, \qquad
    S = (V^-_k)^T BV_k^-,
\end{equation}
that is, for $i=1,\ldots,k$ and $j=1,\ldots,i$,
\begin{equation*}
\begin{aligned}
E_{ij}^\pm & = 
\langle \bm{b}_i^\pm , (A\pm B)\bm{b}_j^\pm \rangle, \\
S_{ij} & = 
\langle \bm{b}_i^- , (\Sigma + \Delta)\bm{b}_j^+ \rangle.
\end{aligned}
\end{equation*}
Once the eigenvalue $\omega_{k+1}$ is computed in the reduced space and the current approximation to the eigenvector $\bm{x}_{k+1}$ has been computed as in eq. \eqref{eq:eigenvector}, 
\begin{equation}
    \label{eq:Ritz}
    \bm{x}_{k+1} = V_{k}(+)\bm{u}^+ + V_{k}(-)\bm{u}^-, \quad \bm{u}^\pm\in\R^{k},
\end{equation}
we can assemble the residual vector in the full space
$R_{k+1}\in\mathbb{R}^{2n}$
\begin{equation}
    \label{eq:res}
    R_{k+1} = (\Lambda - \omega_{k+1}\Omega)\bm{x}_{k+1}, 
\end{equation}
which emerges naturally as the sum of a symmetric term $R_{k+1}(+)$ and an antisymmetric term $R_{k+1}(-)$, whose components are expressed in terms of elements of the spaces $LV_k^\pm$ and $BV_k^\pm$, i.e.,
\begin{equation}
    \label{eq:resp}
    R_{k+1}^\pm = LV^\pm_k \bm{u}^\pm - \omega_{k+1} BV^\pm_k \bm{u}^\mp.
\end{equation}

The expansion subspaces are extended, in the next iteration, using the standard Davidson algorithm, i.e., by solving
\begin{equation}
    \label{eq:DavUpd1}
    (D_\Lambda - \omega_{k+1}D_\Omega)\bm{b}_{k+1} = -R_{k+1},
\end{equation}
where $D_\Lambda$ and $D_\Omega$ are diagonal matrices whose elements are the diagonal elements of $\Lambda$ and $\Omega$, respectively, and $\bm{b}_{k+1} = \bm{b}_{k+1}(+) + \bm{b}_{k+1}(-)$.
Splitting eq. \eqref{eq:DavUpd1} into a symmetric and an antisymmetric part we get, through simple algebra, 
\begin{equation}
    \label{eq:DavUpd3p}
     \widetilde{\bm{b}}_{k+1}^\pm =
     - \left(D_A^2 -\omega_{k+1}^2D_\Sigma^2\right)^{-1}
    \left(D_A R_{k+1}^\pm + \omega_{k+1} D_\Sigma R_{k+1}^\mp\right),
\end{equation}
where $D_A$ and $D_\Sigma$ are the diagonals of $A$ and $\Sigma$, respectively. 
The $\widetilde{\bm{b}}_{k+1}^\pm$ vectors are then orthogonalized to $V_k^\pm$, respectively, and the procedure is iterated until convergence is reached. 
Each iteration requires four matrix-vector products (MVP), to compute $\bm{\sigma}^+_k$, $\bm{\sigma}_k^-$, $\bm{\tau}_k^+$, and $\bm{\tau}_k^-$,
where the first two MVP are typically the cost-dominating step; see eqs. \eqref{eq:MV1}, \eqref{eq:MV2}. However, the cost of solving the generalized eigenvalue problem \eqref{eq:reduced} and the cost of orthogonalizing the new expansion vectors can also become computational bottlenecks, especially if a large number of eigenvalues and eigenvectors are seeked. 
To lower the cost of the former operation, Helmich-Paris \cite{helmich2019casscf} recently proposed a method to reduce the generalized eigenvalue problem into a half-dimensional problem, i.e., the size of one of the blocks $E^\pm$ or $S$. 
The proposed algorithm scales with $\mathcal{O}(k^3)$, where $k$ is the size of the subspace, which is an important improvement with respect to the $\mathcal{O}((2k)^3)$ solution to eq. \eqref{eq:reduced}. However, it requires a sizable number of linear algebra operations, including two singular value decompositions (SVD) and an orthogonalization with respect to a non-positive definite metric, which can become rather expensive. 

In this paper, we propose a different approach that results in a much simpler implementation and that performs, in a $k$-dimensional space, only one diagonalization, one matrix-matrix multiplication, and two Cholesky decompositions. 

We start by noting that, if the ground state solution is stable, problem \eqref{eq:caseigcomp} has no vanishing eigenvalue and therefore it is possible to rewrite it as \cite{olsen1985linear}
\begin{equation}
    \label{eq:asscaseig}
    \Omega \bm{x} = \lambda \Lambda \bm{x}, \qquad 
    \lambda = \frac{1}{\omega}.
\end{equation}
Problem \eqref{eq:asscaseig} is much easier to solve than problem \eqref{eq:caseigcomp}, as it is a generalized eigenvalue problem with a symmetric, positive definite metric $\Lambda$. It is therefore possible to introduce a positive definite dot product
\begin{equation}
    \langle \bm{x}, \bm{y} \rangle_\Lambda = \bm{x}^T \Lambda \bm{y},
\end{equation}
and the induced $\Lambda$-norm $\|\bm{x}\|_\Lambda=\sqrt{\langle \bm{x}, \bm{x} \rangle_\Lambda}$.
We proceed similarly to what has already been presented and expand the eigenvector $\bm{x}$ as the linear combination of symmetric and antisymmetric expansion vectors as in eq. \eqref{eq:eigenvector},
where in our approach we choose the expansion vectors such that
\begin{equation}
    \label{eq:blambda}
    \langle \bm{b}_i^+, \bm{b}_j^+\rangle_\Lambda = \delta_{ij}, \quad 
    \langle \bm{b}_i^-, \bm{b}_j^-\rangle_\Lambda = \delta_{ij},
\end{equation}
that is, the expansion vectors are $\Lambda$-orthogonal.

With the definition of $S$ given in eq. \eqref{eq:redmatE}, we obtain the following reduced problem
\begin{equation}
    \label{eq:redortho}
    \begin{pmatrix}
        0 & S^T \\
        S & 0 \\
    \end{pmatrix}
    \begin{pmatrix}
        \bm{u}^+ \\ \bm{u}^-
    \end{pmatrix}
    = \lambda_{k+1}
    \begin{pmatrix}
        \bm{u}^+ \\ \bm{u}^-
    \end{pmatrix},
\end{equation}
which, due to the choice of $\Lambda$-orthogonal expansion vectors, is no longer a generalized eigenvalue problem. Furthermore, its block structure allows us to solve by computing
\begin{equation}
    \label{eq:halfred}
    S^TS \bm{u}^+ = \lambda^2 \bm{u}^+,
\end{equation}
a symmetric and positive definite eigenvalue problem of size $k$, and then by recovering $\bm{u}^-$ as
\begin{equation}
    \label{eq:u+u-}
    \bm{u}^- = \frac{1}{\lambda} S \bm{u}^+.
\end{equation}
After assembling the current approximation to the eigenvector, we compute the residual
\begin{equation}
    \label{eq:res2}
    R_{k+1} = (\Omega-\lambda_{k+1}\Lambda) \bm{x}_{k+1},
\end{equation}
which can be again split into a symmetric and an antisymmetric term, whose components are
\begin{equation}
    \label{eq:res3}
    R_{k+1}^\pm = BV_k^\pm \bm{u}^\mp - \lambda_{k+1} LV_k^\pm \bm{u}^\pm.
\end{equation}
The new pair of expansion vectors are then obtained by solving
\begin{equation}
    \label{eq:newoth1}
    (D_\Omega - \lambda_{k+1} D_\Lambda)\bm{b}_{k+1} = - R_{k+1}.
\end{equation}
Splitting the new vector and the residual in the sum of symmetric and antisymmetric components, after some computation, we get
\begin{equation}
    \label{eq:DavUpdOrt1}
    \widetilde{\bm{b}}^\pm_{k+1} = -
    \left(\lambda_{k+1}^2 D_A^2 - D_\Sigma^2\right)^{-1}
    \left(\lambda_{k+1}D_A R^\pm_{k+1} + D_\Sigma R^\mp_{k+1}\right).
\end{equation}
The new expansion vectors are then first $\Lambda$-orthogonalized to $V_k^\pm$, and then $\Lambda$-orthonormalized. This is done as follows. First, we project out $V_k^\pm$ from the vectors
\begin{equation}
    \label{eq:bortho1}
    \widehat{\bm{b}}_{k+1}^\pm = \widetilde{\bm{b}}^\pm_{k+1} - V_k^\pm (V_k^\pm)^T\widetilde{\bm{b}}^\pm_{k+1}
\end{equation}
and then we orthonormalize the resulting vectors using the Cholesky decomposition of their overlap, as described in a recent paper by some of us. \cite{nottoli2023robust}
To ensure the numerical stability of this procedure, these two steps are iterated until the norm of $ (V_k^\pm)^T\widetilde{\bm{b}}^\pm_{k+1}$ is smaller than a (tight) threshold. In all our numerical experiments, two iterations were sufficient to obtain vectors orthogonal to machine precision. The resulting vectors $\bar{\bm{b}}_{k+1}^\pm$ are then orthogonalized with respect to $\Lambda$ by computing the Cholesky decomposition of
\begin{equation}
    \label{eq:bortho2}
    M^\pm = (\bar{\bm{b}}_{k+1}^\pm)^T \bm{\sigma}_{k+1}^\pm = L^\pm(L^\pm)^T
\end{equation}
and then solving the triangular linear system
\begin{equation}
    \label{eq:bortho3}
    \bm{b}^\pm_{k+1}(L^\pm)^T = \bar{\bm{b}}_{k+1}^\pm.
\end{equation}
Performing the $\Lambda$-orthogonalization after a regular orthogonalization may look unnecessary, but it vastly improves the conditioning of $M$, thus ensuring the overall numerical stability and robustness of the procedure. Note that, to perform the latter passage, we need to apply $A+B$ and $A-B$ to the new test vectors. In other words, the number of MVP in our algorithm is the same as in the traditional ones, but the linear algebra operations are both simpler and less expensive. In particular, the only $\mathcal{O}(k^3)$ operations that we need to perform are the matrix-matrix multiplication $S^TS$ and the diagonalization in eq. \eqref{eq:halfred}. The price to pay is that we need to perform two more orthogonalizations (eqs. \eqref{eq:bortho2} and \eqref{eq:bortho3}); however, these have a cost of order $\mathcal{O}(2nm^2+m^3)$, $m$ being the number of seeked eigenvectors, and should therefore be less critical.

The overall procedure is summarized in the following algorithm. The algorithms of the primitives used for the orthogonalizations are given in Supporting Information. 
\begin{algorithm}[H]
\caption{\textsc{Generalized Davidson with explicit orthogonalization}\label{GenDav}}
\begin{flushleft}
\textbf{Input:} Initial guess $\bm{x}_0 = (\bm{y}_0, \bm{z}_0)^T$,
procedures to apply $A+B$, $A-B$, $\Sigma+\Delta$, $\Sigma-\Delta$ to a given vector, 
procedure to apply the preconditioner in eq. \eqref{eq:DavUpdOrt1} to a vector, tolerance $\tau$, maximum size of the subspaces $m_{\rm max}$, maximum number of iterations $k_{\rm max}$. Primitives \texttt{b{\_}ortho} and \texttt{b{\_}ortho{\_}vs{\_}x}\\
\textbf{Output:} $\bm{x}$, $\lambda$, the lowest pencils of the generalized eigenvalue problem $\Omega \bm{x} = \lambda \Lambda \bm{x}$.
\end{flushleft}
\begin{algorithmic}[1]
    \State Assemble $\bar{\bm{b}}_1^\pm = \bm{y}_0 \pm \bm{z}_0$
    \State $\bar{\bm{\sigma}}_1^\pm = (A\pm B) \bar{\bm{b}}^\pm_1$
    \State $\bm{b}_1^\pm, \bm{\sigma}^\pm_1$ = \texttt{b{\_}ortho}
    ($\bar{\bm{b}}_1^\pm,\bar{\bm{\sigma}}_1^\pm$)
    \State $k = 0$, $m = 0$
    \State Set $V^\pm_{1} = \bm{b}^\pm_1$, $LV^\pm_1 = \bm{\sigma}^\pm_1$, $BV^{\pm}_0 = \emptyset $
    \While{$k < k_{\rm max}$}
        \State $k = k + 1$, $m = m + 1$
        \State $\bm{\tau}_k^\pm = (\Sigma \mp \Delta) \bm{b}^\mp_k$
        \State Set $BV^\pm_k = [BV^\pm_{k-1}\ \ \bm{\tau}_k^\pm]$
        \State $S = (V^-_k)^T BV_k^-$
        \State Solve $S^TS \bm{u}^+ = \lambda^2 \bm{u}^+$ and get $\lambda_{k+1} = \sqrt{\lambda^2}$
        \State Compute $\bm{u}^- = \frac{1}{\lambda} S \bm{u}^+$
        \State Build the current approximation to the eigenvector
        \State $R^\pm_{k+1} = BV^{\pm}_{k} \bm{u}^\mp - \lambda_{k+1}LV^{\pm}_{k}\bm{u}^{\pm}$
        \State Lock converged eigenvectors, if done, return
        \If{$m < m_{\rm max}$}
            \State $\widetilde{\bm{b}}_{k+1}^\pm$ = \texttt{precnd}($\lambda_{k+1},R^\pm_{k+1}$)
            \State $\bar{\bm{b}}_{k+1}^\pm$ = \texttt{b{\_}ortho{\_}vs{\_}x}($\widetilde{\bm{b}}^\pm_{k+1},V^{\pm}_{k}$)
            \State $\bar{\bm{\sigma}}_{k+1}^\pm = (A\pm B) \bar{\bm{b}}^\pm_{k+1}$
            \State $\bm{b}_{k+1}^\pm, \bm{\sigma}^\pm_{k+1}$ = \texttt{b{\_}ortho}($\bar{\bm{b}}_{k+1}^\pm,\bar{\bm{\sigma}}_{k+1}^\pm$)
            \State Set $V^\pm_{k+1} = [V^\pm_{k} \ \ \bm{b}_{k+1}^\pm]$, $LV^\pm_{k+1} = [LV^\pm_{k} \ \ \bm{\sigma}_{k+1}^\pm]$
        \Else
            \State $m = 1$
            \State Restart Davidson
        \EndIf
    \EndWhile
\end{algorithmic}
\end{algorithm}
Algorithm \eqref{GenDav} is general and can deal with any generalized eigenvalue problem as in eq. \eqref{eq:CASGenEig}. It exhibits monotonic convergence of the reduced eigenvalues, as the Hylleras-Undheim-McDonald theorem applies, \cite{hylleraas1930numerische, macdonald1933successive} and can be implemented efficiently using \textsc{Blas} and \textsc{Lapack} routines.  

\begin{figure}
\includegraphics[width=0.5\textwidth]{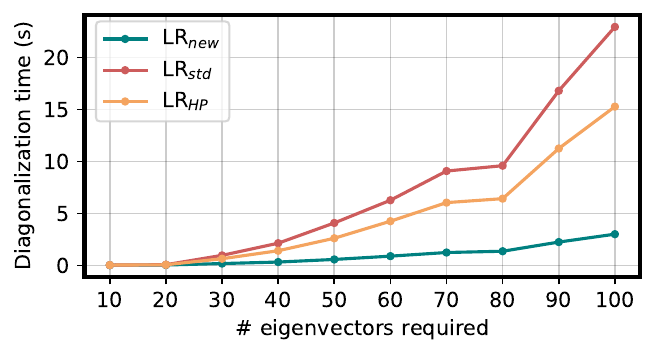}
\caption{\label{fig:time}Time analysis of the three methods discussed for a $2n \times 2n$ problem, where $n = 10000$. Diagonalization time with respect to the number of eigenvalues required.}
\end{figure}

\begin{figure}
\includegraphics[width=0.5\textwidth]{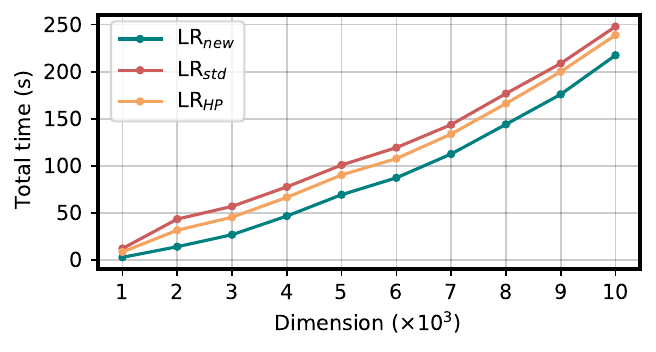}
\caption{Total time (s) of the three methods for different dimensions $n$ and 100 eigenvectors required.\label{fig:scaling}}
\end{figure}

In the following, we report a few numerical tests of the proposed algorithm and compare it with the original algorithm proposed by Olsen et al.\cite{olsen1988solution}, where the $2n\times 2n$ reduced problem is solved, and the algorithm recently proposed by Helmich-Paris\cite{helmich2019casscf}, that uses a series of transformation to reduce the size of the problem to $n\times n$. To compare the three algorithms in a fair way, we implemented them in an in-house code where we generate test matrices with the correct structure and use in-core matrix-vector multiplications to compute the required matrix-vector products. This allows us not only to validate our implementation against dense \textsc{Lapack} routines but also to create common grounds to compare the three algorithms in a fair way. We focus our analysis on the time required to solve the reduced problem and to orthogonalize the new test vectors against the previous ones, as all the other steps are identical.
We generate symmetric and positive definite $A+B$ and $A-B$ matrices by putting
\begin{equation*}
    (A+B)_{ij} = 
    \begin{cases}
      5+i, & \text{if}\ i=j, \\
      \frac{1}{i+j}, & \text{otherwise},
    \end{cases}
\end{equation*}
and
\begin{equation*}
    (A-B)_{ij} = 
    \begin{cases}
      2+i, & \text{if}\ i=j, \\
      \frac{0.2}{i+j}, & \text{otherwise}.
    \end{cases}
\end{equation*}
$A$ and $B$ are then obtained as linear combinations. The symmetric $\Sigma$ and the antisymmetric $\Delta$ matrices are generated in the following way: the former is obtained by generating a random matrix and then multiplying it by its transpose, so to ensure that it is symmetric and positive definite, while the latter antisymmetric matrix is obtained by generating a random matrix and subtracting its transpose.

As a first test example, to investigate the elapsed diagonalization times, we consider a generalized eigenvalue problem of size $2n \times 2n$, where $n = 10000$ is the size of the matrices $A, B, \Sigma$, and $\Delta$. 
We solved the generalized eigenvalue problem by applying the three algorithms discussed in this work: the original procedure proposed by Olsen and coworkers (LR$_{\text{std}}$)\cite{olsen1988solution}, where the reduced problem is solved for $1/\omega$ to have a symmetric, positive-definite metric, the algorithm recently introduced by Helmich-Paris (LR$_{\text{HP}}$) \cite{helmich2019casscf}, and our implementation shown in algorithm \ref{GenDav} (LR$_{\text{new}}$). We seek 10 to 100 eigenpairs with increments of 10. Convergence is achieved when the root-mean-square norm of the residual is smaller than $10^{-6}$ and its maximum absolute value is smaller than $10^{-5}$. We use a subspace dimension of 20, i.e., we keep in memory up to 20 vectors per eigenvector in the history. For all the algorithms we exploit a locking procedure for the converged eigenvectors. 
In figure \ref{fig:time}, we report the cumulative elapsed time required to solve the reduced eigenvalue problem for the three algorithms. Such timings are the sum over all the iterations; note that by design the iterations produced by all algorithms are equivalent, and therefore the number of iterations is shared by all methods.
For the algorithm presented here (LR$_{\rm new}$), the reported timings include the time spent for the additional orthogonalization with respect to the metric (line 20 of algorithm \ref{GenDav}).
We observe that the proposed algorithm significantly outperforms both the original Olsen's algorithm and the new method presented by Helmich-Paris. The differences become more significant as the size of the reduced problem increases, i.e., when more eigenvalues are seeked. Note that, in our simple-minded example, the overall cost is still dominated by the dense matrix-vector multiplications. Nevertheless, the difference in timings is sizeable, and can be expected to have an impact on large-scale applications when many states are computed.
To better appreciate the overall difference between the algorithms, we plot in figure \ref{fig:scaling} the total time required to compute 100 eigenvalues and eigenvectors for increasingly larger systems with $n$ ranging from 1000 to 10000. In all cases, our LR$_{\text{new}}$ algorithm is faster than the others and, in particular, the time difference increases as the dimension decreases, consistently with what was observed in figure \ref{fig:time}. 

In conclusion, we have presented a new algorithm to solve the CASSCF linear response equations that is not only more efficient than what previously reported in the literature, but also conceptually simple and easy to implement.

Thanks to the robust orthogonalization procedures described in the supporting information, it is also numerically robust and stable. If the expansion space becomes ill-conditioned, which is bound to happen near convergence, the metric in the reduced space can exhibit small (numerically zero) or even negative eigenvalues, independent of whether the actual metric is ill-conditioned. This can make the overall procedure fail. By choosing expansion vectors that are orthogonal with respect to the scalar product induced by the metric, we avoid this problem from the beginning. Furthermore, by limiting the linear-algebra operations to a symmetric diagonalization in the subspace, we avoid further propagation of possible numerical instabilities. 
The combination of robustness and efficiency makes therefore the new algorithm an ideal strategy to tackle the solution to the linear response equations.

\begin{acknowledgments}
The authors acknowledge financial support from ICSC-Centro Nazionale di Ricerca in High Performance Computing, Big Data, and Quantum Computing, funded by the European Union -- Next Generation EU -- PNRR, Missione 4 Componente 2 Investimento 1.4.
F.L. and F.P. further acknowledge funding from the Italian Ministry of Research under grant 2020HTSXMA\_002 (PSI-MOVIE).
\end{acknowledgments}


\appendix
\section{Supporting Information}
We report here the routines deployed in the three algorithms to perform dense algebra operations.
All the matrix-matrix products were performed by using level-3 BLAS DGEMM routine.

In Olsen and coworkers' proposed algorithm (LR$_{\text{std}}$),\cite{olsen1988solution}  the $2k\times 2k$ generalized eigenvalue problem in the reduced space is solved by rewriting the equation, as in Eq.14 in the main paper. The equation then takes the form:
\begin{equation*} 
    \label{eq:reduced}
    \begin{pmatrix}
        0 & S^T \\
        S & 0 \\
    \end{pmatrix}
    \begin{pmatrix}
        \bm{u}^+ \\ \bm{u}^-
    \end{pmatrix} =
    \frac{1}{\omega_{k+1}}
    \begin{pmatrix}
        E^+ & 0 \\
        0 & E^- \\
    \end{pmatrix}
    \begin{pmatrix}
        \bm{u}^+ \\ \bm{u}^-
    \end{pmatrix}.
\end{equation*}
This problem is solved by means of LAPACK's DSYGV. The new trial vectors obtained for each iteration (Eq. 13 of the main paper) are orthogonalized with respect to the other vectors in the expansion space by using the routine \texttt{ortho{\_}vs{\_}x}. This routine is a particular case of  \texttt{b\_ortho{\_}vs\_x} -- described in Algorithm \ref{alg:b_ortho_vs_x} -- where the metric $B$ is the identity. 
The orthogonalization of guess trial vectors before the iterative procedure are performed by using the routine \texttt{ortho{\_}cd}, described in Algorithm \ref{alg:ortho_cd}. Both \texttt{ortho{\_}vs{\_}x} and \texttt{ortho{\_}cd} are also described in a recent paper by some of us.\cite{nottoli2023robust}

In Helmich-Paris' work (LR$_{\text{HP}}$), two Singular Value Decompositions (SVD) of the matrix $S$ and auxiliary matrix $C$ (see Ref.\cite{helmich2019casscf}) are performed by using LAPACK's DGESVD, and two Cholesky decompositions of the reduced-space Hessian blocks in the basis of singular vectors are done through LAPACK's DPOTRF (Eqs.18-21 of Ref.\cite{helmich2019casscf}).
Similarly to the LR$_{\text{std}}$ algorithm, orthogonalization procedures of trial vectors are performed by means of \texttt{ortho{\_}cd} and \texttt{ortho{\_}vs{\_}x} (Algorithms \ref{alg:ortho_cd} and \ref{alg:b_ortho_vs_x}).

In our procedure (LR$_{\text{new}}$), the only dense algebra operation is the diagonalization of the matrix $S^TS$ (Eq.14 of the main paper), which is performed by using LAPACK's DSYEV. The orthogonalizations of guess trial vectors and new trial vectors with respect to the metric are performed by using routines \texttt{b{\_}ortho} and \texttt{b{\_}ortho{\_}vs{\_}x}, respectively, described in Algorithm \ref{alg:b_ortho} and Algorithm \ref{alg:b_ortho_vs_x}. A summary of the used dense algebra routines discussed in this section are reported in TABLE \ref{tab:my_label}.  
\begin{algorithm}[H]
\caption{\texttt{ortho\_cd}$(\widetilde X)$: Orthonormalize a set of vectors $\widetilde{X}$ using the Cholesky decomposition of the overlap with iterative refinement.}
\hspace*{\algorithmicindent} \textbf{Input:}  non orthogonal vectors $\widetilde{X}$, threshold $\tau_{\rm ortho}$. \\
\hspace*{\algorithmicindent} \textbf{Output:} $X$, orthonormal vectors.
\label{alg:ortho_cd}
 \begin{algorithmic}[1]
 \State $X = \widetilde X$
 \While{$\|X^{T} X - Id\| > \tau_{\rm ort}$}
     \State $M = {X}^T{X}$
     \State Attempt Cholesky factorization $M = LL^T$
     \If{fail}
         \State Add $\alpha \epsilon \|X\|$ to the diagonal of $M$ until successful
     \EndIf
     \State $X = {X} L^{-T}$
 \EndWhile
 \end{algorithmic}
 \end{algorithm}
\begin{algorithm}[H]
\caption{\texttt{b\_ortho}$(\widetilde X, \widetilde{B X})$: Orthonormalize a set of vectors $\widetilde{X}$ with respect to the metric $B$ using the Cholesky decomposition of the overlap.}
\hspace*{\algorithmicindent} \textbf{Input:}  non orthogonal vectors $\widetilde{X}$, $\widetilde{B X}$, threshold $\tau_{\rm ortho}$. \\
\hspace*{\algorithmicindent} \textbf{Output:} $X$, $B X$ orthonormal vectors.
\label{alg:b_ortho}
 \begin{algorithmic}[1]
   \State $M = {\widetilde X}^T \widetilde{B X}$
   \State Cholesky factorization $M = LL^T$
   \State $X = {\widetilde X} L^{-T}$
   \State $B X = \widetilde{B X} L^{-T}$
 \end{algorithmic}
\end{algorithm}
\begin{algorithm}[H]
 \caption{\texttt{b\_ortho\_vs\_x}$(X, \widetilde Y)$: given a set of $B$-orthonormal vectors $X$ and a set of vectors $\widetilde{Y}$, $B$-orthogonalize $\widetilde{Y}$ to $X$ and orthonormalize $\overline{Y}$.}
 \hspace*{\algorithmicindent} \textbf{Input:}  $B$-orthonormal vectors $X$, non orthogonal vectors $\widetilde{Y}$, threshold $\tau_{\rm ortho}$. \\
\hspace*{\algorithmicindent} \textbf{Output:} orthogonal vectors $\overline{Y}$, $B$-orthogonal with respect to $X$

 \label{alg:b_ortho_vs_x}
 \begin{algorithmic}[1]
 \State $Y = \widetilde{Y}$
 \While{$\|Y^{T}BX\| > \tau_{\rm ortho}$}
     \State $Y = Y - X\left(BX\right)^TY$
     \State $\overline{Y}$ = {\texttt{ortho\_cd}(Y)}
 \EndWhile
 \end{algorithmic}
 \end{algorithm}
\begin{table}[]
    \centering
    \begin{tabular}{lcc}
         \toprule
         Algorithm & diagonalization & orthogonalization \\
         \midrule
         LR$_{\text{std}}$ & DSYGV & 2$\times$ortho\_vs\_x\\
         LR$_{\text{HP}}$ & 2$\times$DGESVD, 2$\times$DPOTRF & 2$\times$ortho\_vs\_x\\
         LR$_{\text{new}}$ & DSYEV &2$\times$b\_ortho\_vs\_x, 2$\times$b\_ortho\\
         \bottomrule
    \end{tabular}
    \caption{Summary of the routines used by the three driver algorithms.}
    \label{tab:my_label}
\end{table}
\newpage


%

\end{document}